\documentclass{article}
\usepackage[accepted]{vietnam} 
\usepackage{natbib}
\usepackage{graphicx}      
\usepackage{hyperref}
\setcitestyle{numbers}

\def\Journal#1#2#3#4{{#1} {\bf #2}, #3 (#4)}

\begin{document}
\twocolumn[
\title{Tracing accretion variability of high-mass YSOs via light echoes}
\titlerunning{Tracing accretion variability of high-mass YSOs via light echoes}
\author{B. Stecklum$^1$, S. Heese, S. Wolf$^2$, A. Caratti o Garatti$^3$, J. M. Iba\~nez$^4$, and H. Linz$^5$}{\newline stecklum@tls-tautenburg.de}

\address{$^1$TLS, Germany; $^2$Kiel University, Germany; $^3$DIAS, Ireland; $^4$CSIC, Spain; $^5$MPIA, Germany}

\keywords{star formation}
\vskip 0.5cm 
]

\begin{abstract}
There is growing evidence for disk-mediated accretion being the dominant mode of star formation across nearly the whole stellar mass spectrum. The stochastic nature of this process has been realized which implies an inherent source variability. It can be traced more easily for low-mass YSOs (LMYSOs) since high-mass YSOs (HMYSOs) are still embedded even when reaching the ZAMS. 
While variable reflection nebulae around LMYSOs were among the earliest signs of star formation, little is known on the variability of scattered light from embedded clusters, the birthplaces of HMYSOs. Since the few most massive stars dominate this emission, their variability is literally reflected in scattered light. Moreover, because of their high luminosity, for a given ambient dust density and source distance, the associated nebulosities are much larger than those of LMYSOs. In this case, the light travel time becomes substantial. So the apparent brightness distribution constitutes a light echo, shaped by both the HMYSO variability history and the spatial distribution of the scattering medium. 
We report on early results of a NIR variability study of HMYSOs associated with Class II methanol masers
which aims at revealing a possible correlation between maser flux density and infrared brightness. Additionally, relevant findings for the eruptive HMYSO S255IR-NIRS3 are presented.
\end{abstract}

\section{Introduction}
The scattering of light by dust or electrons in the interstellar medium allows us to trace past radiation outbursts of cosmic objects by observing their echo which results from different light travel times. Obviously, the travel time for the scattered radiation exceeds that of the radiation propagating along the direct line of sight. The spatial brightness distribution of the echo depends on both the light curve of the outburst and the distribution as well as properties of the surrounding scattering medium\,\cite{Couderc,SUGERMAN}. Light echoes arising from optically thin (single scattering) configurations  can be most easily deciphered. The fast brightness rise and the huge luminosity of supernovae promotes them to be prime originators of light echoes. The light echo not only provides a photometric record but can also be studied spectroscopically to obtain time-resolved spectroscopy of the past outburst. For historic supernova events light echo spectroscopy allowed to pin down the supernova type\,\cite{KRAUSE}. Furthermore, the temporal change of the echo offers the unique possibility to reconstruct the 3D structure of the scattering medium.

While young stellar objects (YSOs) occasionally experience brightness outbursts due strongly enhanced accretion (FUor, EXor events\,\cite{HERBIG,AUDARD}), reports on light echoes from YSOs are scarce\,\cite{Ortiz,Muze}. Historically, an outburst of R CrA observed by E. Hubble\,\cite{Hubble} was probably the first YSO light echo observation ever although it was not recognized as such at that time. Recent theoretical results point to disk instabilities as the reason for variable accretion\,\cite{VOROB,MEY,MRI}, and observational findings support the view that there is a continuum of accretion burst behavior\,\cite{KOSPAL,PEÑA,K2}.

In case of YSOs both the circumstellar disk as well as the envelope surrounding YSOs introduce an anisotropic radiation field that governs the spatial morphology of the light echo. For HMYSOs the circumstellar disk is the crucial ingredient which allows ongoing accretion at large luminosities that would have stopped spherical accretion otherwise because of radiation pressure\,\cite{SONNYORKE}. The channeling of the radiation along the YSO symmetry axis as well as the shielding of the matter reservoir by the disk is acknowledged 
as the key mechanism to form massive stars via disk accretion, e.g.\,\cite{KUIPER}.

Class {\sc II} 6.7\,GHz methanol masers, thought to be pumped by thermal IR emission\,\cite{Sobolev}, trace embedded luminous YSOs\,\cite{Breen}. These masers are generally variable, and some show regular flux changes with periods of several 10...100 days\,\cite{Goedhart,Szymczak}. There are competing  models for what concerns the origin of the  periodicity\,\cite{vanderWalt,Inayoshi}.  If the maser variability is caused by modulation of the pumping radiation due to variable accretion or protostellar pulsations then similar changes in the scattered light are expected. Thus,
surface brightness variations of the scattered emission from HMYSOs should be traceable. If this claim holds these brightness variations are expected to be in sync with the methanol maser activity.

\section{Data Retrieval, Processing, and Results}
In order to test this idea work started to analyze Ks-band brightness variations of  a few periodic maser sources which are covered by the VVV survey\,\cite{MINNITI}. Compared to other archival data this survey provides the best observational coverage with up to 50 epochs of NIR imaging which boosts the detection probability for light echoes. One serendipitous discovery in a massive star forming region was reported (D. Minniti, unpublished). 

Rather than performing aperture photometry on the images, a direct detection of surface brightness variations was attempted using a differential image analysis (DIAS) method\,\cite{BRAMICH}. For this purpose, the image with the best seeing (smallest PSF width) was subtracted from all the other ones after convolving it with a kernel to match  the individual PSFs.
\begin{figure*}
\centering
$\begin{array}{cc}
\includegraphics[angle=0,width=15.cm]{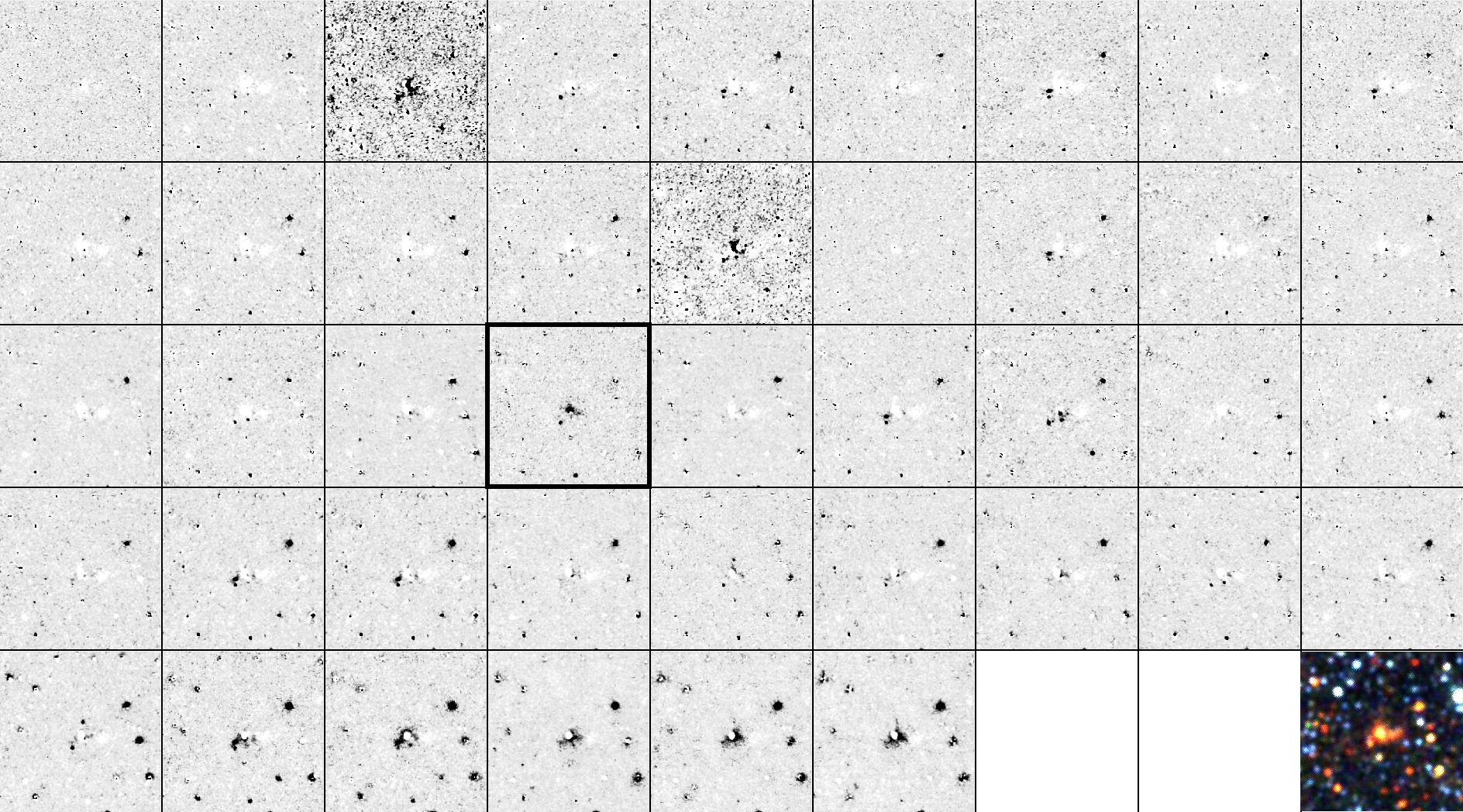} 
\end{array}$
\vskip -0.25cm
\caption{Inverted gray scale DIAS images of 43 VVV Ks frames of the G339.62-00.12 region, sorted according to seeing. The field of view amounts to  about 1.5'$\times$1.5' and is aligned to Galactic coordinates. The lower right insert shows the color composite from 2MASS with the maser source at the center. The solid black lines highlight the frame with a clear emission excess of the maser source.
}
\label{fig:G339}
\end{figure*}
Results for the maser source G339.62-00.12 are shown in Fig.\,\ref{fig:G339}. The image displays the DIAS frames in the order of the seeing FWHM (best one upper left). While most of the frames do not show an excess of scattered emission from the location of the HMYSO, there is at least one clear detection (bold frame at [4,3] position). It has yet to be checked to which phase of the maser light curve\,\cite{Goedhart} the epoch of the particular image corresponds.

 It can also be seen that the DIAS method has problems in matching frames with relatively poor seeing (lowermost row). Nevertheless, it has become obvious that high cadence NIR surveys have the potential to trace brightness variability of deeply embedded HMYSOs through scattered light. With more epochs to come from VVV and its successor VVVX, the prospects for establishing a relation between the maser and NIR variability are improving.

\section{The Outburst of S255IR-NIRS3}
During the works on the VVV data a flare of the Class {\sc II} methanol maser in S255IR was reported\,\cite{FUJISAWA}. Having in mind that the  IR pumping of the maser might be due to an accretion outburst, NIR imaging of this field was sought immediately. In late November 2015, i.e. two weeks after the flare announcement, the first epoch of Ks-band imaging with the PANIC camera\,\cite{CITE} was obtained at the 2.2-m telescope at Calar Alto observatory. This image revealed that the deeply embedded HMYSO S255IR-NIRS3 experienced a brightness outburst of $\sim$2.5\,mag which was also obvious from the enhanced surface brightness of the associated bipolar nebula.  Therefore, follow-up imaging was performed on several occasions throughout 2016. S255IR-NIRS3 is a well-studied HMYSO at a distance of 1.78$\pm$0.11\,kpc\,\cite{Burns} with a luminosity in quiescence of $L\sim 2 \times 10^4\,L_{\odot}$ and a mass of $\sim20\,M_{\odot}$\,\cite{Simpson} which consists of a disk/outflow system\,\cite{Boley,Zinchenko}. These findings suggested a causal relationship between the accretion outburst and the maser flare\,\cite{ATEL}.  

\begin{figure*}[h]
\vskip -0.25cm
\centering
$\begin{array}{cc}
\includegraphics[angle=0,width=7.cm]{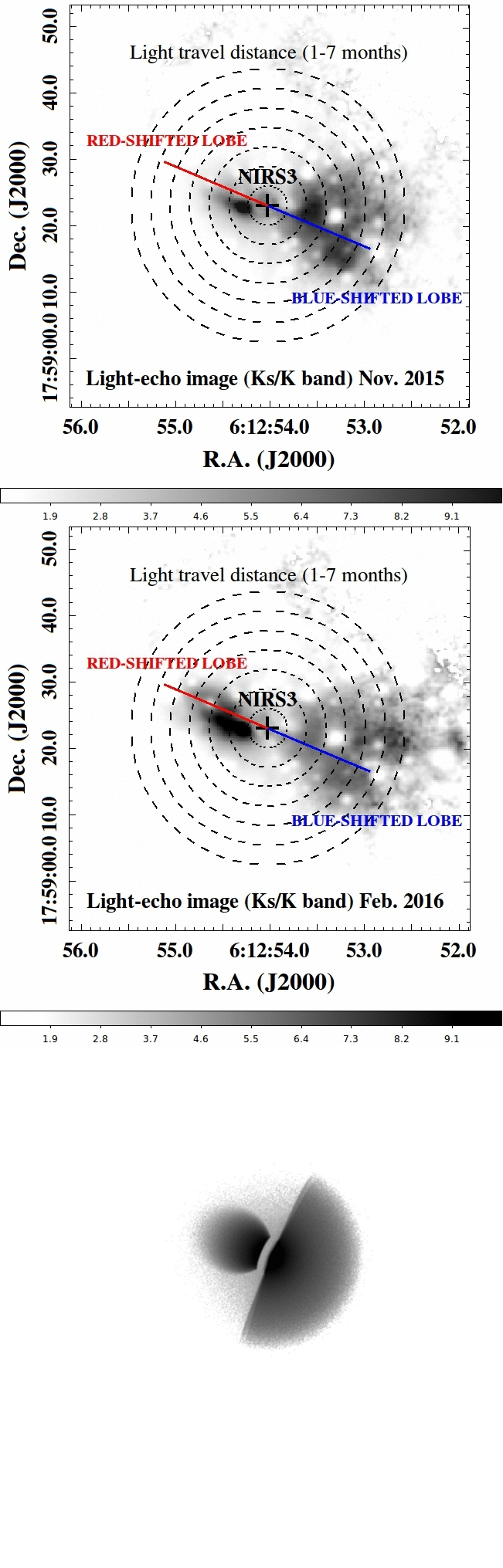} 
\end{array}$
\vskip -0.25cm
\caption{Top and center - brightness ratio images showing the bipolar light echo and its propagation. The gray-scale bar at the bottom indicates the ratio values. Bottom - Echo simulation for a circumstellar disk with an inclination of 75$^{\circ}$.
}
\label{fig:two}
\end{figure*}

In order to ease the assessment of the surface brightness increase, the PANIC images were divided by an archival pre-burst UKIDSS K-band image (epoch 2009) after matching the PSF widths. Thereby, spatial extinction variations across the field as well as the geometric dilution of the radiation cancel. The resulting ratio images (cf. Fig.\,\ref{fig:two}) clearly show enhanced scattering with a striking bipolar morphology. While the enhanced scattering could be caused by an increase of the volume density of dust grains this explanation is not applicable since there is no mechanism which could establish such an increase within six years. So the conclusion could be drawn that the enhanced scattering represents the light echo of the burst. It was verified by tracing the motion of the echo in the 2015 Nov and 2016 Feb images, respectively. By averaging the extent of the blue- and red-shifted echoes an approximate onset date of the burst could be established which points to 2015 June.

In order to extract the intrinsic burst light curve from the echo observations a proper scattering model is required. To this aim the Monte Carlo radiative transfer code Mol3D\,\cite{Ober} which includes multiple scattering was modified by incorporating time tags for each synthetic photon. Fig.\,\ref{fig:two}(bottom) shows a first result of the modeling along with the two brightness ratio images. The model image shows the response of a circumstellar disk embedded in a uniform density medium to an instant stellar flash of infinitesimal duration. It verified that the code is working properly but is not meant to already reproduce the case of NIRS3.

While the NIR spectrum of NIRS3 is almost featureless, spectroscopy of the scattered light before (archival data) and during the burst revealed spectral features which provide additional evidence for the accretion burst from S255IR-NIRS3\,\cite{ALESSIO}.

\section{Summary}
Preliminary results of studying temporal variations of scattered light from HMYSOs based on VVV data indicate that surveys with sufficient cadence offer a great potential to reveal variability of massive young stars which cannot be seen in direct light because of high extinction.  Thus it seems promising to apply this method to investigate the relationship between the variability of HMYSOs and their associated Class {\sc II} methanol masers, in particular for what concerns periodic behavior. The case of S255IR-NIRS3 showed that accretion bursts from HMYSOs can trigger flares of those masers. The analysis of the light echo allowed to estimate the time delay between onset of the burst and the maser flare which amount to about two months. By reverting the case it can be speculated that methanol maser variability might be an indirect tracer of variable accretion of massive young stars.

\section*{Acknowledgments}
B.S. acknowledges support by the DFG grant STE 605/27-1.

\end{document}